\documentclass[preprint,12pt]{elsarticle}

\usepackage{amssymb}

\journal{Nuclear Physics B}

\usepackage[capitalize]{cleveref}

\begin{document}
    
    \begin{frontmatter}
        
        
        
        \title{Conformal Mapping Approach to Dipole Shim Design}
        
        
        \author{Thomas Planche\footnote{tplanche@triumf.ca}\,, Paul M.\ Jung, Suresh Saminathan, Rick Baartman}
        \address{TRIUMF, 4004 Wesbrook Mall, Vancouver, BC, Canada V6T 2A3}
        \author{Matthew J.\ Basso\footnote{Now at the University of Toronto}}
        \address{University of British Columbia Okanagan, 3333 University Way, Kelowna, BC, Canada V1V 1V7}
        \date{\today}
        
        \begin{abstract}
            Passive shims are often used to reduce the size and cost of room-temperature magnetic dipoles. In this paper we revisit an analytic approach to the problem of optimum shim design, and we extend it by taking into consideration the effect of magnetic saturation. We derive an abacus curve to determine optimum shim dimensions as a function of the desired dipole nominal field. 
            We show that, for nominal fields below 1.2\,T, a pole with such shims can be made at least one half gap height narrower than a pole without.
            We discuss the range of validity of this approach and verify its predictions using 2 and 3-dimensional finite-element calculations. 
            
        \end{abstract}
        
        
    \end{frontmatter}
    
    
    \section{Introduction}
    In the context of accelerator magnet design, a shim is a device used to provide fine adjustments of the magnetic field profile. 
    There are two types: active shims, which are extra coils mounted on the poles; and passive shims, which  are localized modifications of the pole shape.  
    Active shims are often used in NMR and MRI instruments~\cite{anderson1961electrical,hoult1994accurate,terada2011development} and superconducting particle accelerator magnets~\cite{ferracin2014magnet}.  
    Passive shims are often used in room temperature magnets for particle accelerators~\cite{rose1938magnetic,danby1999precision}, and are generally designed based on empirical rules and  iterative processes involving 2 and/or 3-dimensional finite element calculations~\cite{russenschuck2011shim}.
    To facilitate this design process it is useful to start from a `good guess'. Such initial guess can be chosen among the set of optimum shim dimensions analytically derived by Rose~\cite{rose1938magnetic} using conformal maps. The issue is that the solution derived by Rose is not unique, leaving some room for arbitrariness in the choice of the initial guess. In this paper, after a brief review of Rose's work, we attempt to complete his work by taking into consideration the effect of magnetic saturation.  We show that, for a given nominal field, the effect of magnetic saturation reduces the set of optimum ``Rose'' shims to a narrow range of optimum solutions. We provide an abacus curve to determine optimum shim dimensions as a function of the desired dipole nominal field, and we test the validity of our approach using 2 and 3-dimensional finite-element calculations.

    \section{Magnetic Dipole Shim}\label{sec:2}

    The relative permeability of ferromagnetic materials, which is much larger than 1 at low excitation, converges asymptotically to 1 at high excitation: this is what is referred to as ``magnetic saturation''. 
    As a starting point we make the assumption that the pole surface is not saturated, i.e.\ the pole surface is equipotential. We will discuss the range of validity of this assumption in~\cref{sec:saturation}. We also assume that we work with dipoles which are long compared to their gap height. 
    Under these assumptions, our problem reduces to solving Laplace's equation in 2 dimensions. 
    We also assume that we work with dipoles that are wide compared to their gap height, which allows us to consider only one pole edge at a time. We will show in~\cref{sec:narrowpole} that in fact this approximation remains valid for pole widths down to  about only 1 gap height.
    The last assumption we make is to neglect the direct contribution from the coil to the magnetic field distribution. 
    Note that the direct contribution from the coil could be taken into account following Ref.~\cite{walstrom2012dipole}. But for the sake of simplicity, and as we concern ourselves with optimizing field flatness well inside the magnet gap, we choose to neglect this effect.

    \subsection{Conformal Mapping}
    Conformal mapping can be used to find analytical 2-dimensional solutions to Laplace's equation. The technique depends on constructing a transformation which preserves angles locally, from a geometry for which a solution is known, to the geometry of interest. Detailed introductions to the technique can be found in several text books (see for instance~\cite{weber1950electromagnetic}).
    
\begin{figure}[htb]
        \centering
        \includegraphics*[width=0.61\textwidth]{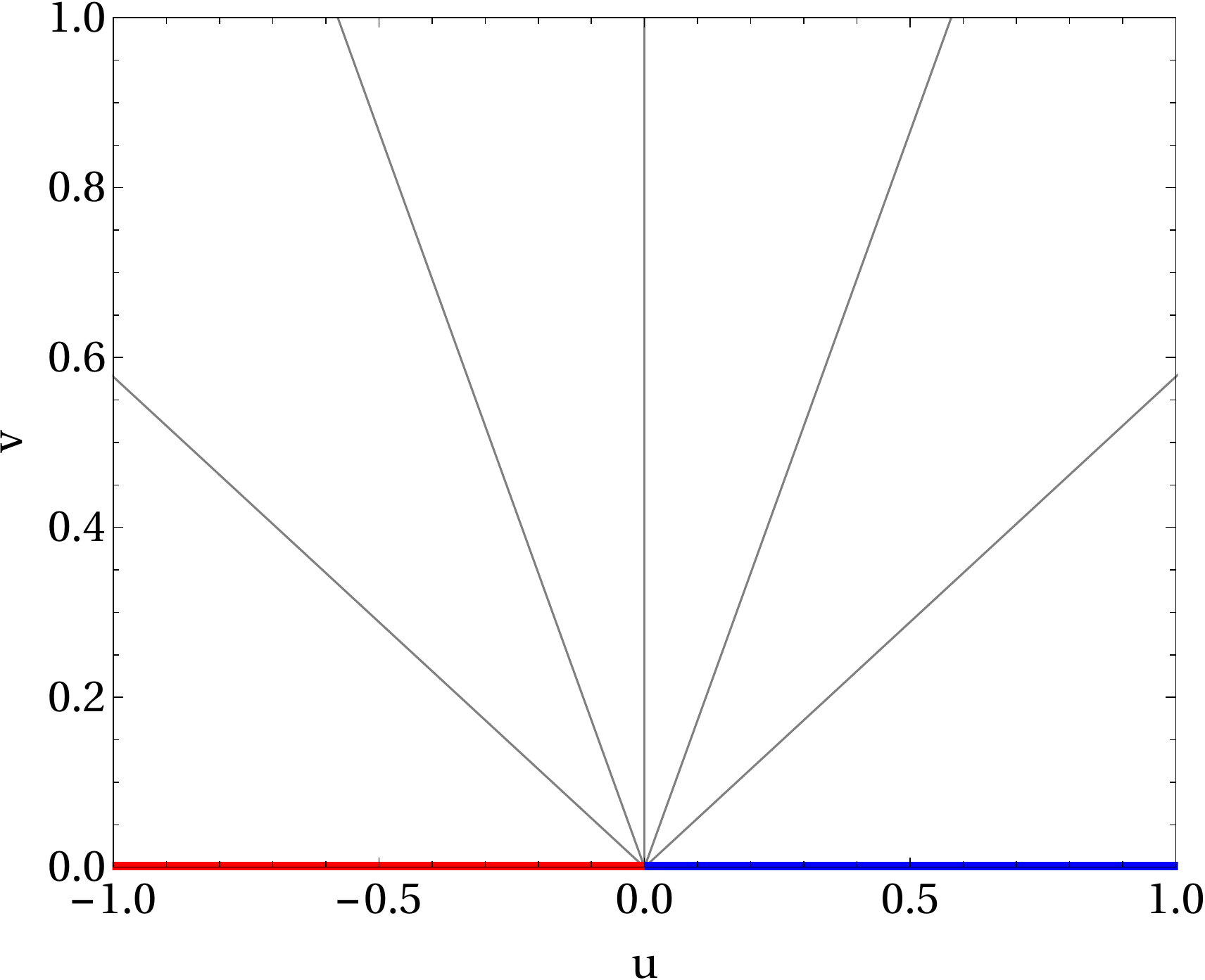}
        \caption{Schematic representation of a flat condenser. The red and blue lines represent constant potential boundaries; they extend to $u=-\infty$, and $u=+\infty$, respectively. A few more equipotential are shown as thin grey lines.}\label{fig:flatcondenser-equipot}
    \end{figure}

    All the geometries presented in this paper are solved starting from the solution of Laplace's equation for a flat condenser (see~\cref{fig:flatcondenser-equipot}):
    \begin{equation}\label{eq:pot}
        \phi(u,v)=\frac{\Delta V }{\pi }\arctan\left(\frac{u}{v}\right) +V_0\,,
    \end{equation}
    where $u$ and $v$ are real.
    One can verify that~\cref{eq:pot} satisfies $\nabla^2 \phi=0$, and that the boundary condition: 
    \begin{eqnarray}
        \phi(u,v=0) & = & \left\{ \begin{array}{lc}
            V_0+\frac{\Delta V }{2 }, & u>0,\\[0.5em]
            V_0-\frac{\Delta V }{2 }, & u<0,\\[0.5em]
            V_0, & u=0.
        \end{array}\right.
    \end{eqnarray}
    is satisfied. 
    In the electro-static case $\Delta V$ is the difference of potential between the plates of the condenser. In the magnetic case $\Delta V=\mu_0 NI$, with $NI$ the amount of current flowing perpendicularly to the ($u,v$) plane along (0,0).
    
    Let $x$ and $y$ be the Cartesian coordinates of the geometry we are trying to solve for.
    If we construct two complex variables: $t=u+iv$ and $z=x+iy$, the relation between them is given by the Schwarz-Christoffel formula:
    \begin{equation}\label{eq:Schwarz}
        \frac{{\rm d}z(t)}{{\rm d}t}=C\prod_n(t-t_n)^{\alpha_n/\pi-1}\,,
    \end{equation}
    where $\alpha_n$ is the internal angle of the $n^{{\rm th}}$ corner of the studied geometry, and  $t_n$ is the value of $t$ associated with this corner in the in complex t-plane. The value of the constant $C$ is determined according to the specific scale of the problem.
    The magnetic field distribution is obtained from:
    \begin{equation}\label{eq:field}
        H(t)=H_x-iH_y=-\frac{{\rm d}\phi}{{\rm d}z}=-\frac{{\rm d}\phi}{{\rm d}t}\left(\frac{{\rm d}z}{{\rm d}t}\right)^{-1}\,.
    \end{equation}
    This way, as noted by Rose~\cite{rose1938magnetic}, the field distribution on the midplane can be obtained as a function of $t$ without analytic integration of~\cref{eq:Schwarz} for complex values of $t$ and $z$.

    \subsection{Rose's Shim}\label{sec:Rose}
\begin{figure}[htb]
        \centering
        \includegraphics*[width=0.61\textwidth]{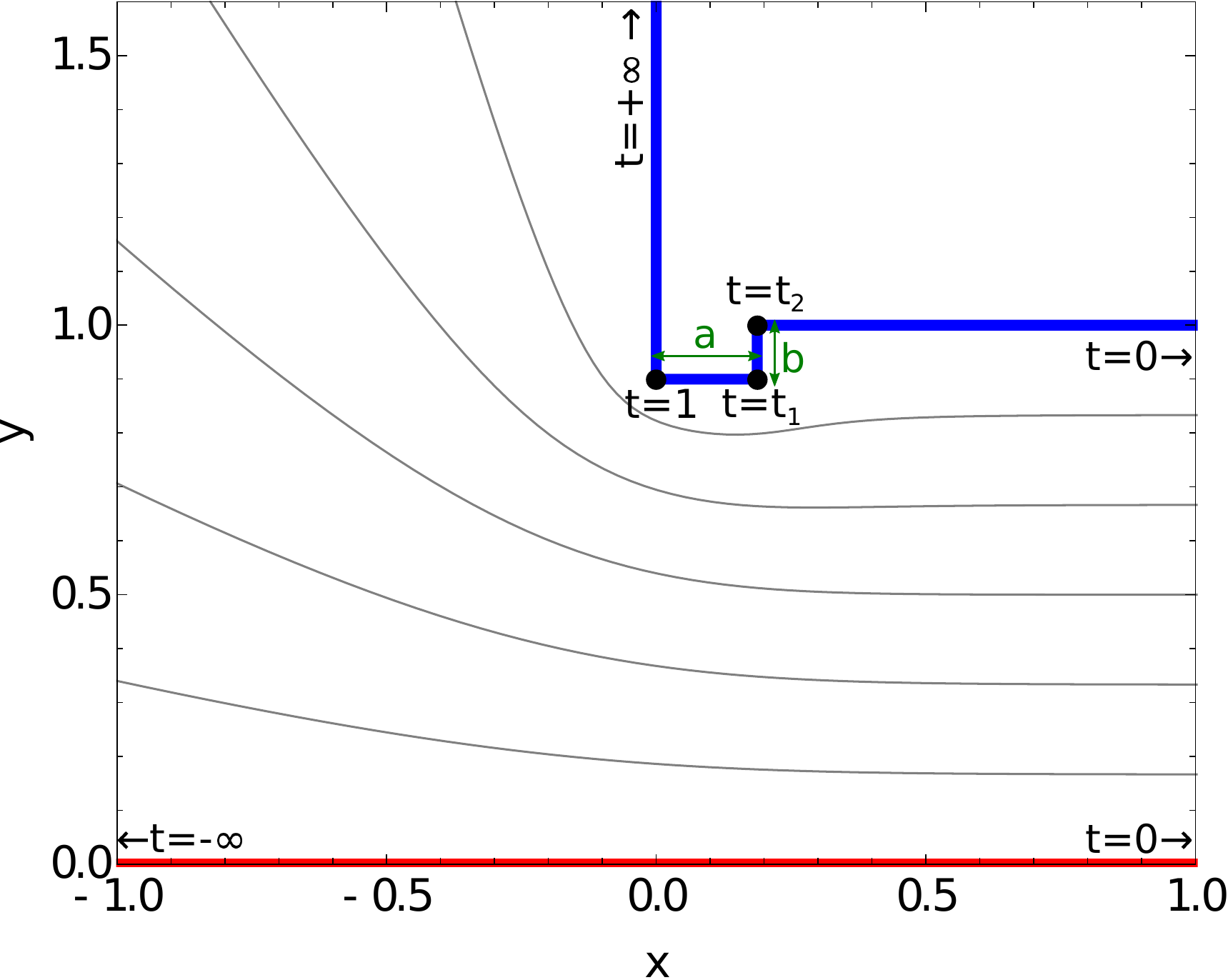}
        \caption{Dipole edge with a rectangular shim. The blue line is the equipotential that goes along the pole surface; the red line is the equipoltential that goes along the magnet mid-plane. A few more equipotentials, obtained from numerical integration of~\cref{eq:zRose} along the equipotential lines of~\cref{fig:flatcondenser-equipot}, are shown as thin grey lines.}\label{fig:Rose}
    \end{figure}

    In this section we revisit Rose's work on rectangular shim design~\cite{rose1938magnetic}.
    Unlike Rose, we choose to take advantage of the up-down symmetry by setting a fixed potential along the dipole mid-plane, see~\cref{fig:Rose}.
    This geometry possesses four corners (see~\cref{tab:Rose}) leading to:
    \begin{equation}\label{eq:zRose}
        \frac{{\rm d}z}{{\rm d}t}=C\frac{ \sqrt{t-1} \sqrt{t-t_1}}{t \sqrt{t-t_2}}\,.
    \end{equation}
    \begin{table}[htb]
        \begin{center}
            \begin{tabular}{ c  c  c  c  }
                \hline
                $n$ & $t_n$ & $\alpha_n$ & $z$ \\ \hline
                0  & 1 & $\frac{3\pi}{2}$ & $i(1-b)$ \\
                1  & $t_1$ & $\frac{3\pi}{2}$ & $a+i(1-b)$ \\
                2  & $t_2$ & $\frac{\pi}{2}$ & $a+i$ \\
                3  & 0 & 0 & $+\infty$\\ \hline
            \end{tabular}
        \end{center}
        \caption{Parameters of the four corners of the Rose shim geometry.}\label{tab:Rose}
    \end{table}
    
\begin{figure}[htb]
        \centering
        \includegraphics*[width=0.7\textwidth]{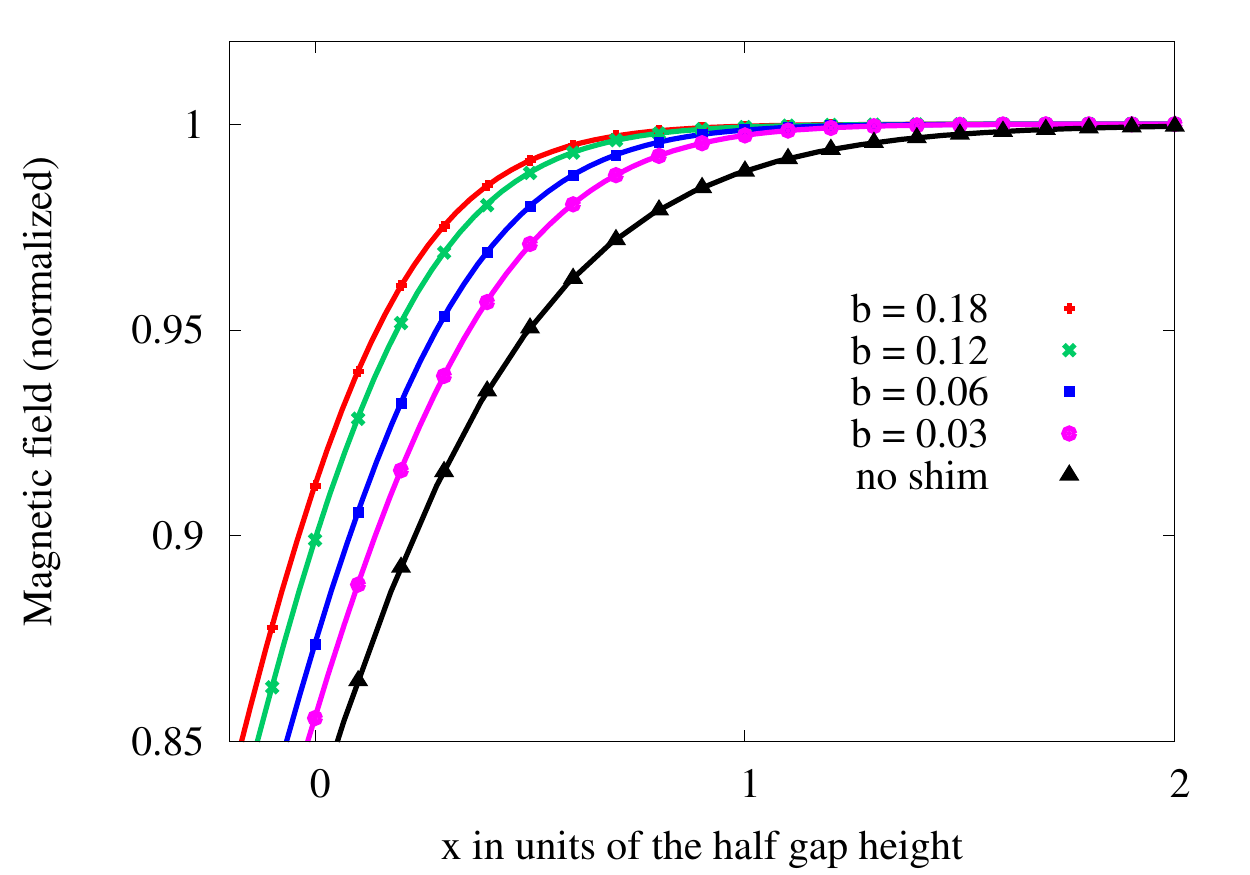}
        \caption{Field distribution along the magnet mid-plane for Rose shims of various heights $b$. Lines are calculated using~\cref{eq:Hy(x)}; round dots are obtained from {\tt Opera-2d} calculations for the same geometry, assuming linear steel properties.}\label{fig:shimeff}
    \end{figure}

    We set the values of $\Delta V$ and $C$ such that both the half-gap height and the magnitude of the field at $t=0$ are equal to 1, leading to:  $\Delta V = -1$, $C =\frac{i \sqrt{t_2}}{\pi \sqrt{t_1}}$, and
    \begin{equation}\label{eq:RoseH}
        H(t)=\frac{\sqrt{t_1 (t-t_2)}}{\sqrt{t-1} \sqrt{t_2
                (t-t_1)}}\,.
    \end{equation}
    At this point the geometry possesses two degrees of freedom: the value of $t_1$ and $t_2$ which determine the value of $a$ and $b$ through:
    \begin{equation}\label{eq:ab}
        a=\int_1^{t_1} \frac{{\rm d}z}{{\rm d}t}{\rm d}t\,,
        \quad  
        b=\int_{t_1}^{t_2} \frac{{\rm d}z}{{\rm d}t}{\rm d}t\,.
    \end{equation}
    The problem consists in finding the particular values of $a$ and $b$ that lead to optimal field flatness inside the magnet. 
    To solve this problem Rose~\cite{rose1938magnetic} expands~\cref{eq:RoseH} around $t=0$:
    \begin{equation}\label{eq:Hexp}
        i\,H(t)=1+\frac{t}{2}  \left(\frac{1}{t_1}-\frac{1}{t_2}+1\right)+O\left(t^2\right)\,.
    \end{equation}
    Expanding~\cref{eq:zRose} and solving this differential equation for $t$ close to 0 leads to:
    \begin{equation}\label{eq:tapprox}
        t(z)\approx \left(i\sin(\pi y)-\cos(\pi y)\right)e^{-\pi x}\,.
    \end{equation}
    Substituting in~\cref{eq:Hexp} leads to 
    an approximate expression of the vertical field along the magnet mid-plane:
    \begin{equation}\label{eq:Hyexp}
        H_y(x)\approx 1-\frac{e^{-\pi x}}{2}  \left(\frac{1}{t_1}-\frac{1}{t_2}+1\right)+O\left(e^{-2\pi x}\right)\,.
    \end{equation}    
    To make the field ``optimally'' homogeneous Rose chooses the value of $t_2$ such that the first order term in $e^{-\pi x}$ vanishes:
    \begin{equation}\label{eq:t2}
        t_2 = \frac{t_1}{1 + t_1}\,.
    \end{equation}
    With this additional constraint the geometry has only one degree of freedom left. 
    This last degree of freedom cannot be used to cancel the next order term as~\cref{eq:Hyexp} now reads:
    \begin{equation}
        H_y(x)\approx 1-\frac{e^{-2\pi x}}{2 t_1} +O\left(e^{-3\pi x}\right)\,,
    \end{equation}
    and $0<t_1<1$. The contribution from the term in $e^{-2\pi x}$ is minimized for shims that are relatively narrow ($a$ small) and tall ($b$ large), see~\cref{fig:shimeff}. 
    The magnetic field distribution along the magnet mid-plane is obtained as a function of the (real) negative parameter $u$ :
    \begin{equation}\label{eq:Hy(x)}
        \left\{\begin{array}{rl}
            H_y(u)&=\Im\{H(u)\}=-\sqrt{\frac{u^2}{(u-1) (t_1-u)}+1}\\
            x(u)&=x_0+\int_{-1}^u  \frac{{\rm d}z}{{\rm d}t} {\rm d}t\\
        \end{array}\right. \,,
    \end{equation}
    where: 
    \begin{equation}\label{eq:x0}
        x_0=z(t=-1)=\Re\left\{\int_{1}^{-1} \frac{{\rm d}z}{{\rm d}t} {\rm d}t\right\} \,.
    \end{equation} 
    Mid-plane field profile obtained from \cref{eq:Hy(x)} and from a 2D finite element calculation using {\tt Opera-2d}~\cite{opera2010opera} show excellent agreement, see~\cref{fig:shimeff}. {\tt Opera-2d} calculations are done here assuming linear steel properties: discussion of the effects of saturation is postponed to~\cref{sec:saturation}.
    
    To provide a rule of thumb for how effective Rose shims are, let us consider a required relative field flatness of a few $10^{-3}$ or better. With this assumption one can see in~\cref{fig:shimeff} that a shim height of 0.03 half gap enables a reduction of the pole width by about 1 half gap. A taller shims can lead to a further reduction of the pole width by another 0.5 half gap.
    
\begin{figure}[htb]
        \centering
        \includegraphics*[width=0.7\textwidth]{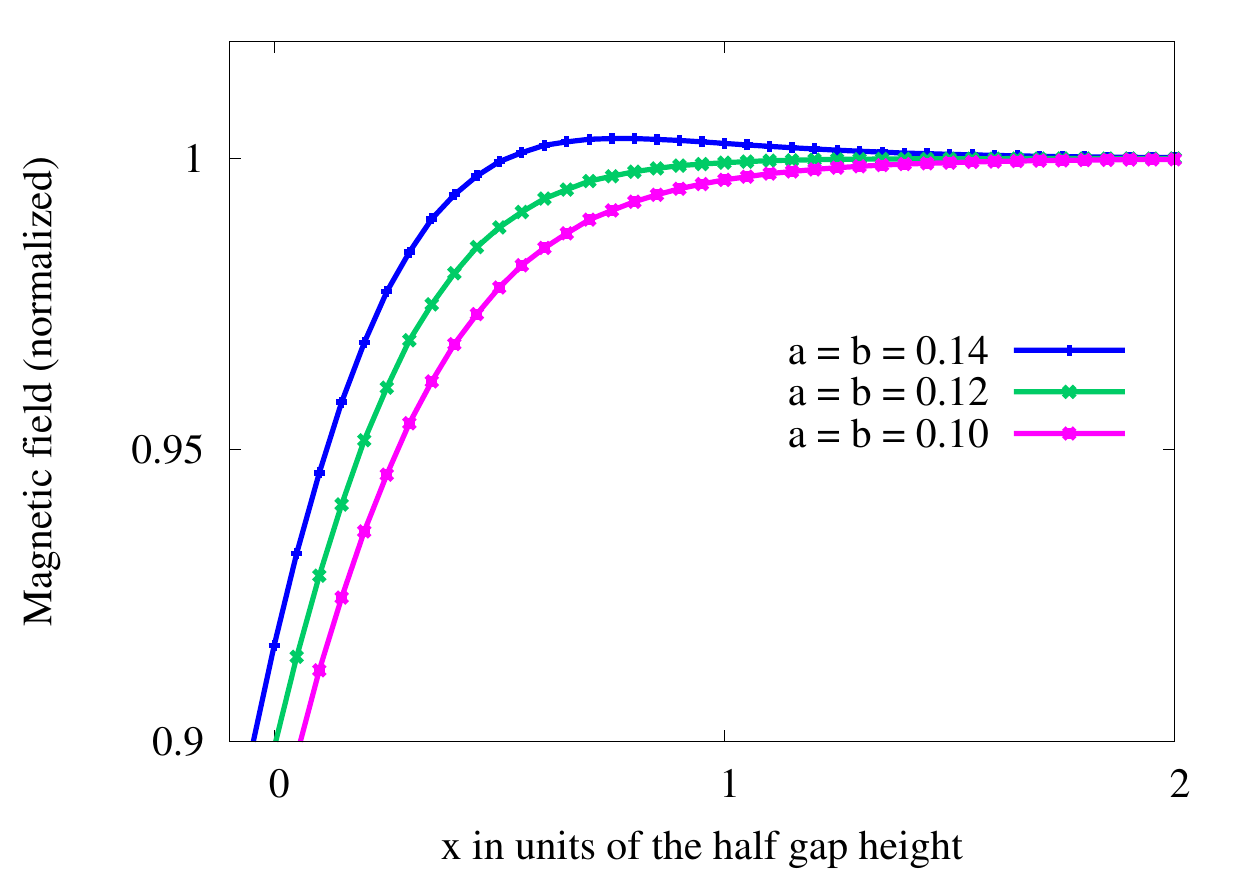}
        \caption{Rose shim ($a=b\approx12\%$) compared to two square shims with non-optimal dimensions.}\label{fig:optimum}
    \end{figure}

    As illustrated in~\cref{fig:optimum}, deviation from the optimum produces a field distribution along the mid-plane which:
    (1) is closer to the no-shim case for cases with ($a$,$b$) below the optimum curve, and (2) overshoots for cases with ($a$,$b$) above the optimum curve.

    \subsection{Shims on Narrow Poles}\label{sec:narrowpole}
\begin{figure}[htb]
        \centering
        \includegraphics*[width=0.7\textwidth]{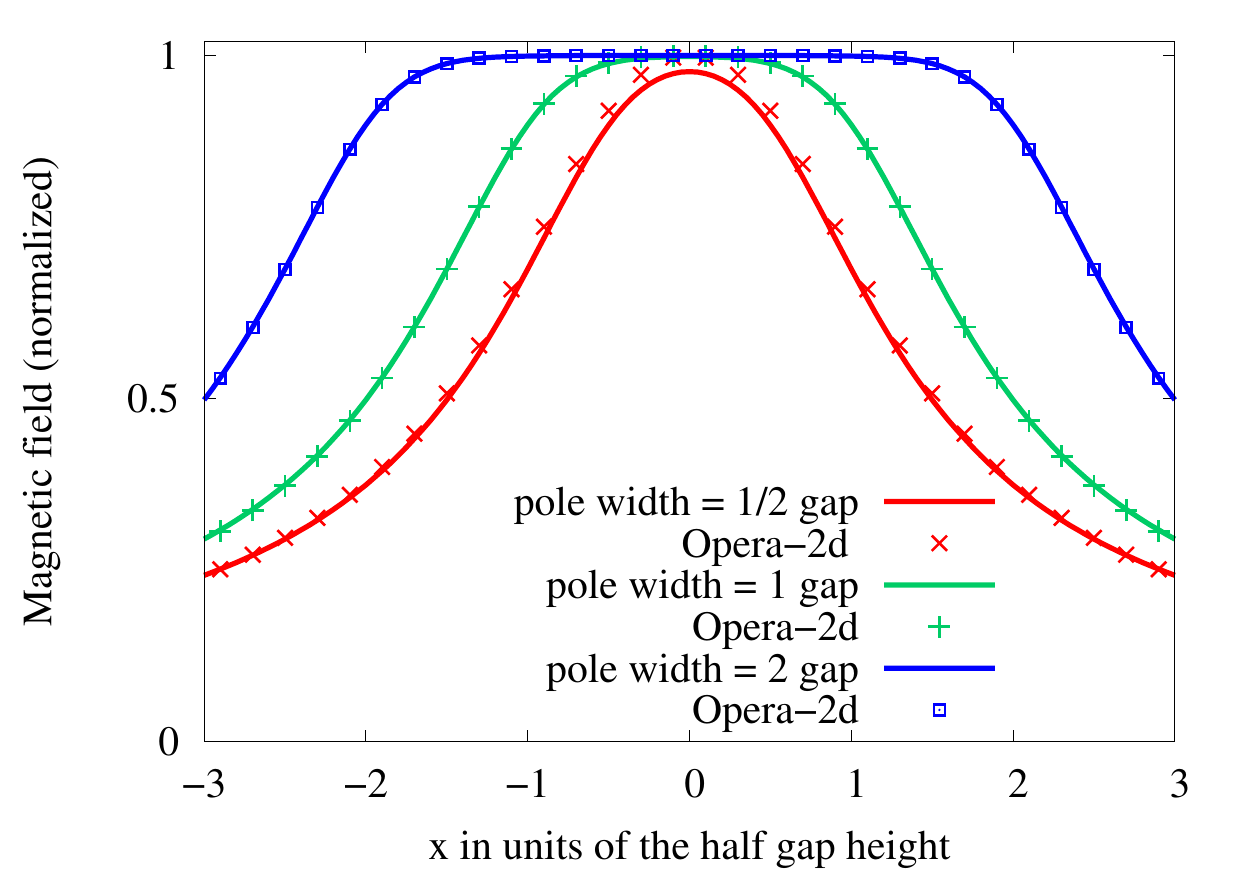}
        \caption{Field distribution along the mid-plane of finite-width poles with Rose shims on both ends ($a=b=0.121$ in unit of half gap). Solid lines are obtained from~\cref{eq:narrow}; points are obtained from {\tt Opera-2d} calculations. }\label{fig:narrow}
    \end{figure}

    The pole geometry used so far, which includes only one edge, is anticipated to be accurate only if the pole is ``wide enough''. We show in~\cref{fig:narrow} that this model remains accurate for pole width down to only about 1 gap height, 
    provided that the contribution from the two edge are superimposed using:
    \begin{equation}\label{eq:narrow}
        \mathcal{H}_y(x)=H_y(x+\frac{w}{2}) + H_y(-x+\frac{w}{2}) - B_0\,,
    \end{equation}    
    where $w$ is the pole width, $H_y(x)$ is obtained from~\cref{eq:Hy(x)}, and $B_0$ is the nominal field (equal to 1 in the units used here).  

    \subsection{Magnetic Saturation}\label{sec:saturation}
    As the magnetic field in the shim approaches saturation, the surface of the material is no longer  equipotential rendering the solution obtained from conformal mapping inaccurate and weakening the effect of the shim.
    When designing magnets, one must choose the height of the shim such that it does not saturate. To guide this choice we will determine the relation between the shim height and the field level inside the shim.

    \begin{figure}[h]
        \centering
        \includegraphics[width=0.7\textwidth]{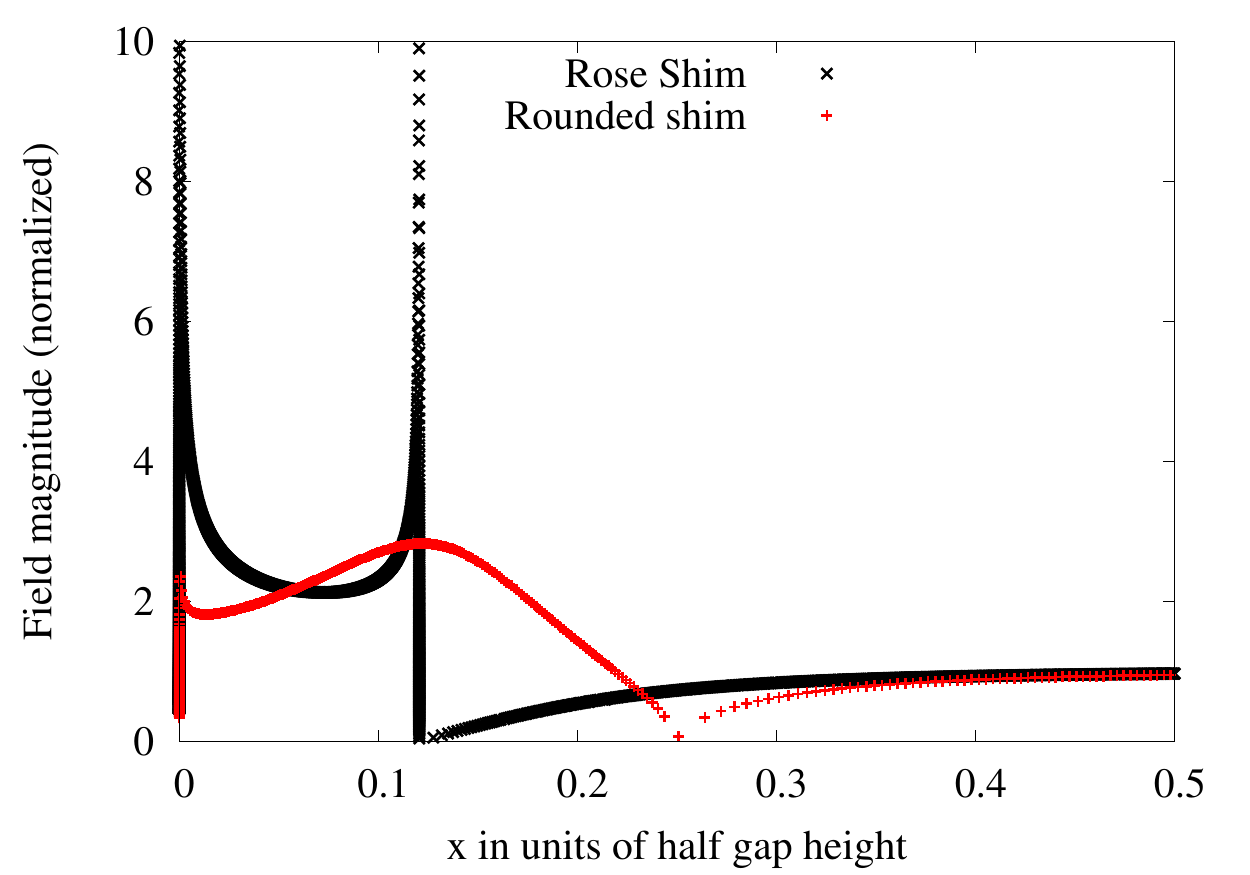}
        \caption{
            Magnitude of the magnetic field along the surface of two types of shim. Black dotted line: a Rose shim with
            $b=a=0.121$; Red solid line: a rounded shim with $b=0.121$ and $a=0.274$. The magnitude of the field is given in units of the magnet nominal field.}\label{fig:RoseSurface}
    \end{figure}

    Conservation of the magnetic flux imposes continuity of the magnitude of the magnetic field across the material surface. This allows us to use the magnitude of the field along the shim surface as a measure of the field level inside the shim.
    The field along the surface is obtained from ~\cref{eq:RoseH} for $t\in\mathbb{R}^+$.
    
\begin{figure}[htb]
        \centering
        \includegraphics[width=0.61\textwidth]{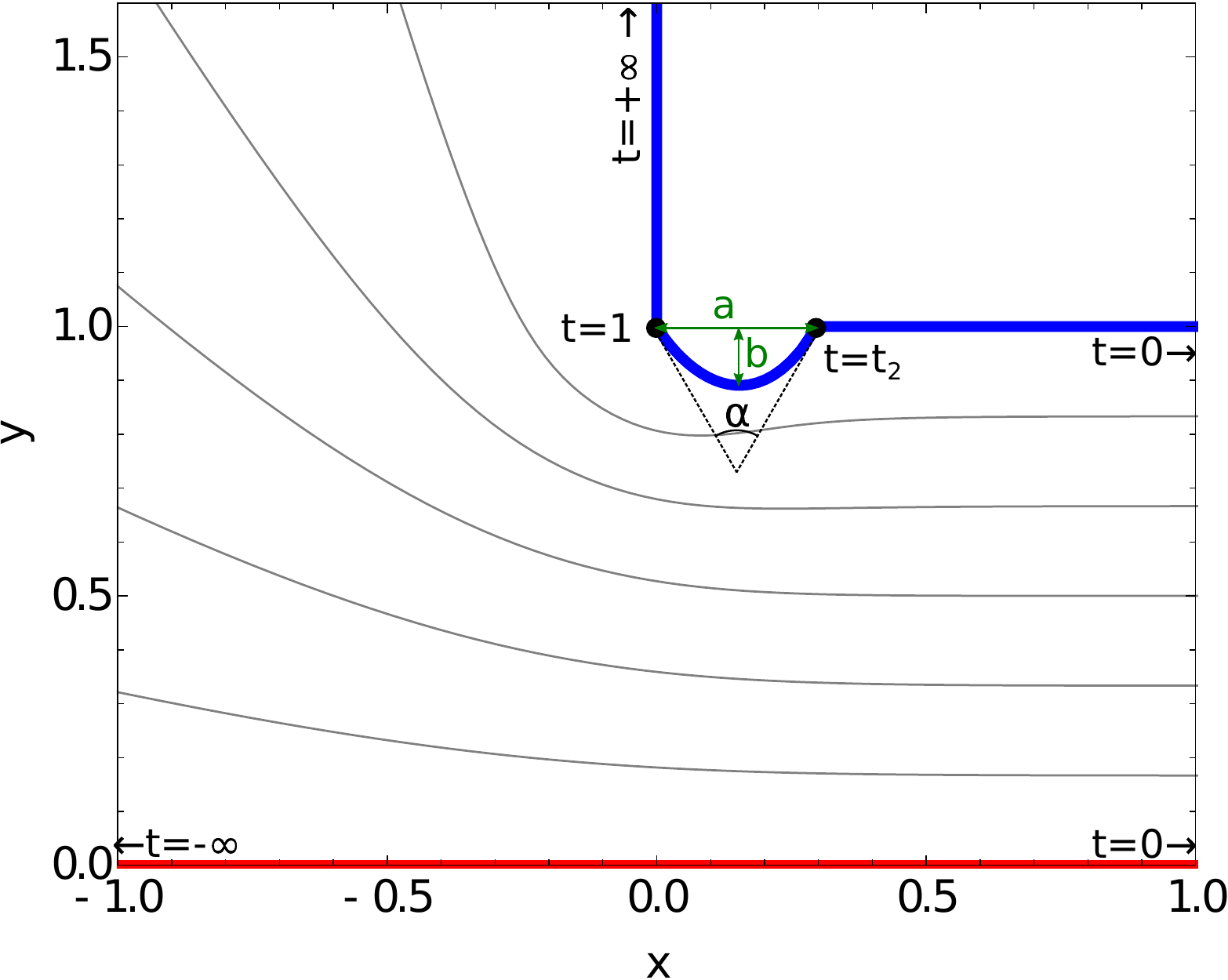}
        \caption{Proposed rounded shim geometry. The effective width $a$ of the shim is controlled by the value of the angle $\alpha$; the height $b$ depends on $\alpha$ and $\lambda$, see~\cref{eq:zisoRound}. This plot corresponds to $\alpha=\frac{\pi}{3}$.}\label{fig:isoRound}
    \end{figure}

    In the case of a rectangular shim the magnitude of the field diverges around its two prominent corners (see~\cref{fig:RoseSurface}) making it hard to discuss field levels inside the steel. To eliminate this singularity we choose to work with a ``rounded'' version of the Rose shim presented in~\cref{fig:isoRound,tab:isoRound}.
    This geometry possesses three corners leading to:
    \begin{equation}\label{eq:zisoRound}
        \frac{{\rm d}z}{{\rm d}t}=\frac{1}{t}C (t-1)^{\frac{\alpha }{2 \pi }}  \left((t-1)^{1-\frac{\alpha }{\pi }}+\lambda 
        (t-t_2)^{1-\frac{\alpha }{\pi }}\right)(t-t_2)^{\frac{\alpha }{2 \pi
            }-\frac{1}{2}}\,,
    \end{equation}
    where the term $(t-1)^{1-\frac{\alpha }{\pi }}+\lambda 
    (t-t_2)^{1-\frac{\alpha }{\pi }}$ is used to produce a corner rounded with a parabola (see~Ref.\cite{weber1950electromagnetic}).
    
    Like in~\cref{sec:Rose} we choose the scale of our problem such that both the half-gap height and the magnitude of the field at $t=0$ are equal to 1, leading to:
    \begin{equation}\label{eq:isoRoundH}
        H(t)=\frac{(t-1)^{\frac{\alpha }{2 \pi }} \left(\frac{t}{t_2}-1\right)^{\frac{\alpha
                    +\pi }{2 \pi }} \left(t_2^{\frac{\alpha }{\pi }}+\lambda 
            t_2\right)}{\lambda  (t-1)^{\frac{\alpha }{\pi }} (t-t_2)+(t-1)
            (t-t_2)^{\frac{\alpha }{\pi }}}.
    \end{equation}
    Once again expanding~\cref{eq:isoRoundH} around $t=0$ and cancelling the lowest order dependence in $t$ leads to:
    \begin{equation}
        \lambda=-\frac{t_2^{\frac{\alpha }{\pi }-1} (\alpha -\alpha  t_2+2 \pi  t_2-\pi
            )}{-\alpha +\alpha  t_2+\pi }\,.
    \end{equation}
    Finally, the value of $t_2$ is set such that the shim starts and ends at the same $y$ value by numerically solving:
    \begin{equation}
        \Im\left\{\int_1^{t_2} \frac{{\rm d}z}{{\rm d}t}{\rm d}t\right\}=0\,.
    \end{equation}
    The only free parameter left is the value of $\alpha$. The shim width $a$ and height $b$ are obtained numerically from:
    \begin{equation}\label{eq:ab2}
        a=\int_{1}^{t_2} \frac{{\rm d}z}{{\rm d}t}{\rm d}t\,,\quad b=\Im\left\{\int_{1}^{t_{\rm tip}} \frac{{\rm d}z}{{\rm d}t}{\rm d}t\right\}\,,
    \end{equation}
    where $t_{\rm tip}$ is the value of $t$ that maximizes $\Im\left\{\int_{1}^{t} \frac{{\rm d}z}{{\rm d}t}{\rm d}t\right\}$ for $t_2<t<1$.
    
    \begin{table}[h!]
        \begin{center}
            \begin{tabular}{ c  c  c  c  }
                \hline
                $n$ & $t_n$ & $\alpha_n$ & $z$ \\ \hline
                0  & 1 & $\pi+\frac{\alpha}{2}$ & $i$ \\
                1  & $t_2$ & $\frac{\pi}{2}+\frac{\alpha}{2}$ & $a+i$ \\
                2  & 0 & 0 & $+\infty$\\ \hline
            \end{tabular}
        \end{center}
        \caption{Parameters of the proposed rounded shim geometry.}\label{tab:isoRound}
    \end{table}

\begin{figure}[htb]
        \centering
        \includegraphics[width=0.81\textwidth]{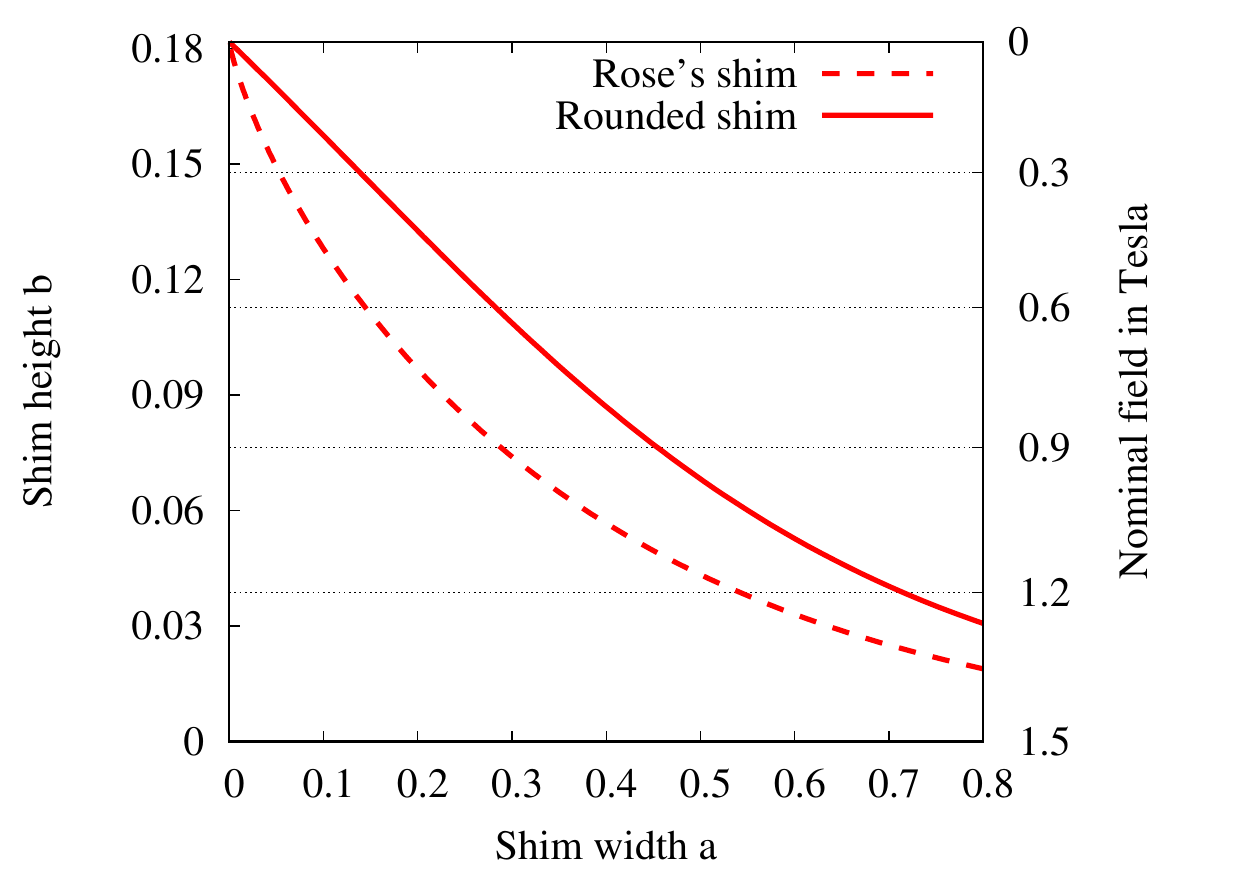}
        \caption{Abacus: optimum effective shim height $b$ as a function of the shim width $a$, given in units of the half gap height, for rectangular (Rose) and rounded shims. The left-hand $y$-axis gives the relation between the magnet nominal field and the  shim height to produce a field of 1.5\,T on the tip of a rounded shim.}\label{fig:abac}
    \end{figure}

    As expected the field along the surface of the rounded shim is continuous, see~\cref{fig:RoseSurface}.
    We now know how to calculate the magnitude of the magnetic field at the tip of the shim as a function of the shim height.
    In~\cref{fig:abac} we present a plot similar to Fig.~3 of Ref.~\cite{rose1938magnetic}: it gives the optimum shim height as a function of the shim width, given in units of the half-gap height. This essential addition to this figure is the second $y$-axis: it gives the relation between the magnet nominal field and the maximum shim height, assuming a saturation field of 1.5\,T. 
    
    \section{Example of application}\label{sec:bonnie}
    The authors were tasked with modifying an existing dipole magnet, not for beam transport, but to serve as a test stand to study penning discharges. 
    The magnet half gap is 238.8\,mm, and the required nominal field is 0.3\,T. The existing pole is narrower in the longitudinal ($z$) direction than in the transverse ($x$) direction. To improve field uniformity we considered installing shims on the edges of the pole in the $z$ direction.
    
\begin{figure}[htb]
        \centering
        \includegraphics[width=0.6\textwidth]{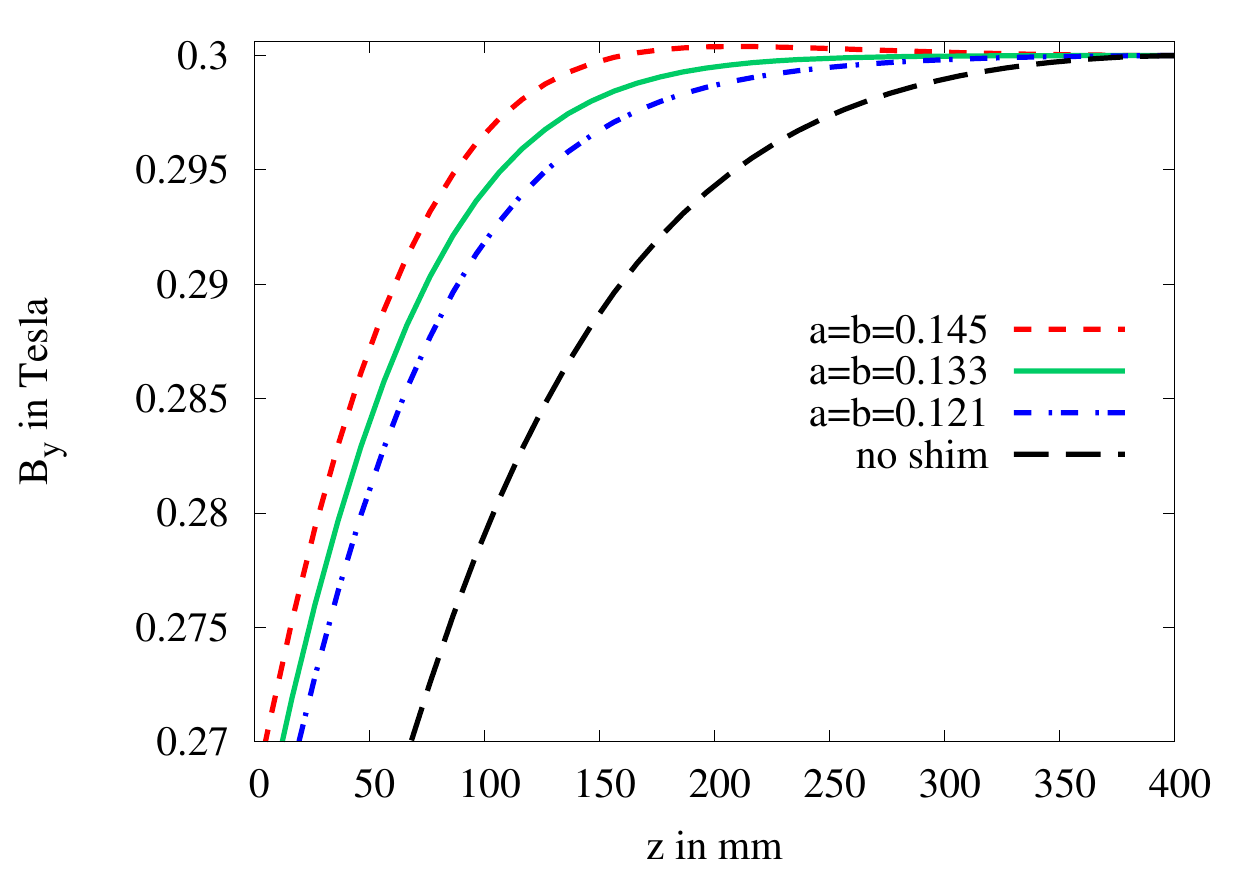}
        \caption{Effect of a square shim ($a=b$) calculated using {\tt Opera-3D} along the $z$-axis with $x=y=0$. The edge of the magnet is at $z=0$, and the magnet center is at $z=400$\,mm. The optimum square shim size is found to be $a=b=0.133$. Note that with a larger shim ($a=b=0.145$) the field slightly overshoots.}\label{fig:bonnie}
    \end{figure}

\begin{figure}[h!]
        \centering
        \includegraphics[width=0.6\textwidth]{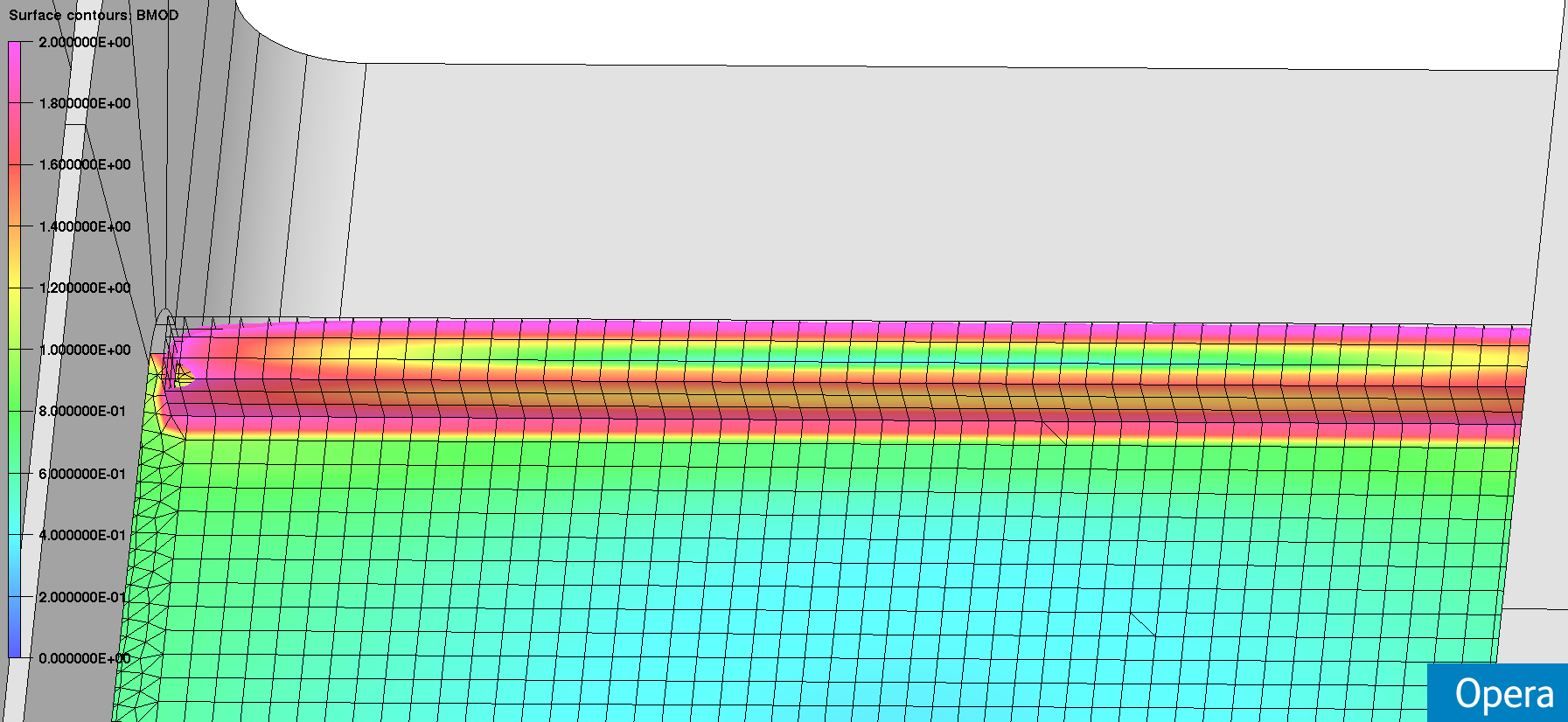}
        \caption{\label{fig:iso} Isometric projections of the {\tt Opera-3d} model. Shim height and width are both equal to 13.3\% of the half-gap height. The colour scale shows the magnitude of the magnetic field on the surface of the steel, dark blue: 0\,T, green: 0.8\,T, orange: 1.5\,T, purple: 2.0\,T, transparent over 2.0\,T. The coil is in grey. Thin black lines materialize the finite-element mesh. Non-linear magnetic properties of the steel are modelled using the B(H) curve of a typical C1010 steel.  
        }
    \end{figure}

    In~\cref{fig:abac} we see that with a nominal field of 0.3\,T we can use a shim height up to about 14.5\% of the half gap. We choose to attempt to use a square shim ($a=b$). In~\cref{fig:abac} the line $a=b$ crosses the Rose shim line at $a=b=12.1\%$, and it crosses the rounded shim line at $a=b=14.5\%$. This defined the narrow range within which to choose our initial guess. We implemented the square shim geometry into {\tt Opera-3d}, solved the optimization problem, and found that the optimum square shim dimension is around $a=b=13.3\%$, hence half-way between the two lines in~\cref{fig:abac}. The corresponding field flatness is presented in~\cref{fig:bonnie}.
    As predicted, \cref{fig:iso} shows that, although the corners are saturated, the tip of the shim is below 1.5\,T.
    
    \section{Conclusion}
    Given the nominal field of an iron dominated electro-magnet, we have shown how to determine the optimum dimensions of simple passive shims (see~\cref{fig:abac}). 
    Passive shims are useful to reduce the size of magnetic and electro-static dipoles. 
    As a rule of thumb we have shown that for a nominal field below 1.2\,T one can use a shim height of at least 0.03 half gap, which leads to a pole width reduction of at least 1 half gap.
    
    In this paper, we restricted ourselves to the study of magnetic dipoles. All 
    results however can be generalized to quadrupoles and any higher-order multipoles by means of an extra conformal transformation~\cite{halbach1968application}. These results can also be applied to electro-static elements, for which the surface field limit ensues, not from a saturation effect, but from the Kilpatrick limit.
    
    \bibliographystyle{elsarticle-num}
    \bibliography{Mybib,AllDN}

    %
    %
    %
\end{document}